\documentclass[12pt]{article}
 
\usepackage{epsfig,amssymb}
\begin{document}

\newcommand{\be}{\begin{equation}}
\newcommand{\ee}{\end{equation}}
\newcommand{\bea}{\begin{eqnarray}}
\newcommand{\eea}{\end{eqnarray}}
\newcommand{\da}{\dagger}
\newcommand{\dg}[1]{\mbox{${#1}^{\dagger}$}}
\newcommand{\hlf}{\mbox{$1\over2$}}
\newcommand{\lfrac}[2]{\mbox{${#1}\over{#2}$}}
\newcommand{\scsz}[1]{\mbox{\scriptsize ${#1}$}}
\newcommand{\tsz}[1]{\mbox{\tiny ${#1}$}}
\newcommand{\gogo}{{\bf ~(*** ACTION ITEM ***)~}} 
\newcommand{\godo}{{\bf ~(*** NOTE ***)~}} 
\newcommand{\goend}{{\bf ~(*** NOTE END ***)~}} 


\begin{flushright}
astro-ph/0501626 \\
LA-UR-04-9081 \\
\end{flushright}  


\begin{center}
\vspace{0.165in}

\Large{\bf Directly Measured Limit on the Interplanetary \\ 
Matter Density from Pioneer 10 and 11}

\vspace{0.4in}

\normalsize 

{\bf Michael Martin Nieto,${^a}$ Slava G. Turyshev,$^b$ and John D. Anderson$^b$} \\

\vspace{0.3in}

{\it
${^a}$Theoretical Division (MS-B285), Los Alamos National Laboratory,\\
University of California,  Los Alamos, New Mexico 87545, U.S.A.} \\
E-mail: mmn@lanl.gov         

\vskip 15pt

{\it
$^{b}$Jet Propulsion Laboratory, California Institute of  Technology,\\
Pasadena, CA 91109, U.S.A.} \\ 
Email: turyshev@jpl.nasa.gov, john.d.anderson@jpl.nasa.gov

\end{center}

\baselineskip=.33in

\begin{abstract}
The Pioneer 10 and 11 spacecraft had exceptional deep-space navigational capabilities. The accuracies of their orbit reconstruction were limited, however, by a small, anomalous, Doppler frequency drift that can be interpreted as an acceleration of $(8.74\pm1.33)\times 10^{-8}$ cm/s$^2$ directed toward the Sun.  We investigate the possibility that this anomaly could be due to a drag on the spacecraft from their passing through the interplanetary medium.  Although this mechanism is an appealing one, the existing Pioneer radiometric data would require an 
unexpectedly high mass density of interplanetary dust for this mechanism to work. Further, the magnitude of the density would have to be nearly constant at 
distances $\sim$20-70~AU.  Therefore, 
it appears that such an explanation is very unlikely, if not ruled out. 
Despite this, the measured frequency drift 
by itself places a {\it directly-measured, model-independent} limit of $\lesssim 3 \times 10^{-19}~\mathrm{g/cm}^3$ on the mass density of interplanetary dust in the outer ($\sim$20-70~AU) solar system.  Lower experimental limits can be placed if one presumes a model that varies with distance. An example is the limit $\lesssim 6 \times 10^{-20}~\mathrm{g/cm}^3$ obtained for the model with an axially-symmetric density distribution that falls off as the inverse of the 
distance.  We emphasize that the limits obtained are {\it experimentally-measured, 
in situ} limits.  A mission to investigate the anomaly would be able to place a 
better limit on the density, or perhaps even to measure it.

\end{abstract}

\begin{center}
\today
\end{center}



\newpage

\section{Introduction}

Due to their long distances from the Sun, their spin-stabilized attitude control, and their long, continuous, radio-tracking Doppler data histories, 
very precise orbit reconstructions could be obtained for 
Pioneer 10 and 11.  Because of this, the Pioneers were very sensitive detectors for a number of solar system effects; in fact, much more sensitive than any other spacecraft in deep space \cite{pioprl,pioprd}. However, despite their excellent navigational capabilities, the accuracies of the Pioneer orbit reconstructions were limited by a small anomalous, constant, one-way Doppler frequency drift 
of size $(5.99 \pm 0.01) \times 10^{-9}$ Hz/s.\footnote{
Depending on the particular piece of data and the fitting procedure, the exact size and formal error vary slightly. (This particular number comes from the ``experimental" result for $a_P$ determined with Pioneer 10 data as described in Section VI of Ref. \cite{pioprd} and is referenced to the downlink carrier frequency, 2.29 GHz.)  
But these differences are much less that the size of the anomaly and also significantly less than the systematic error.} 
This frequency drift is clearly 
present in the data from both Pioneer craft.

Three separate analyses using independent orbit determination codes 
have confirmed the presence of this anomalous frequency drift in the radiometric Doppler data received from the Pioneer 10 and 11 spacecraft when they were at large heliocentric distances, $\sim$ 20-70  AU 
\cite{pioprl,pioprd,markward}.\footnote{\label{footdistance}
The precisely analyzed Pioneer 10 data was taken between 3 January 1987 and 22 July 1998 (when the craft was 40 AU to 70.5 AU distant from the Sun) 
while that from Pioneer 11 was obtained between 
5 January 1987 and 1 October 1990 (22.4 to 31.7 AU) \cite{pioprd}.  
Earlier, not thoroughly analyzed data seems to indicate the anomaly may exist 
as close in as 10 AU from the Sun.  (See Figure 7 of \cite{pioprd}.)  
The goal is to analyze this data in the future.}  
The detected effect can be interpreted as an acceleration $a_{\tt P}= (8.74 \pm 1.33) \times 10^{-8}$  cm/s$^2$ acting in the approximate  direction  {\it towards}  the Sun \cite{pioprl,pioprd}. This interpretation has became known as the Pioneer anomaly.  

A possible mechanism to explain the anomaly is that the craft experience a ``drag'' from their passing through the interplanetary medium 
\cite{pioprd,courtens,foot,foot2,piofind}.\footnote{
While the Pioneer Doppler data was being investigated 
in Refs. \cite{pioprl,pioprd}, this possibility was considered. 
Indeed, one of the motivations to look at the data from the Galileo and Ulysses spacecraft was the possibility that their multifrequency tracking capabilities 
would allow any ``drag'' signal for the origin of the Pioneer anomaly to be seen. 
Unfortunately, because of  individual engineering problems with these craft, the results were inconclusive. (For more details see Ref. \cite{pioprd}.)}  
Just as for a sail boat, where the relative momentum of the air to the craft's sail determines the force, so too for a spacecraft.  The relative velocity of the dust (and, with the dust's mass, its momentum) with respect to the cross-sectional 
area of the spacecraft determines the force on the craft.

Recently, discoveries of many extra-solar planets with unexpected properties 
(such as major planets moving in orbits that come close to their parent stars 
\cite{newplanets}) suggest that the formation of planetary systems 
may be significantly different than previously believed. Further, 
infrared observations have found high dust densities around many main 
sequence stars \cite{irstellardust}.   (The caveat is that the highest 
densities tend to be around younger, brighter, and more massive stars.)   
Even so, and together with our relatively limited knowledge of the outer 
parts of our own solar system, these discoveries reopen the question of whether 
there exists as yet undiscovered interplanetary dust in this distant region. 

This raises the possibility that momentum transfer between the dust distribution and the moving Pioneers could, in principle,  
provide enough power to slow the spacecraft down at a nearly constant rate.  
Here we investigate if a drag force could indeed be the origin of the Pioneer anomaly. In Section \ref{sec2} we explain the physics of the drag mechanism  and review our knowledge of the interplanetary medium in Section \ref{sec3}. We find, in Section \ref{sec4}, that an unexpectedly large amount of interplanetary dust would be needed to cause the anomaly.  
Even so, our result yields a new {\it in situ, experimental, model-independent} limit on the mass density of the medium, since this (expected to be lower) density has not been measured, only modeled.  We give our conclusions and an overview of future work in Section \ref{sec5}.


\section{The Pioneer anomaly as a drag force}
\label{sec2}

\begin{figure}[h]
\begin{center}\noindent    
\epsfig{figure=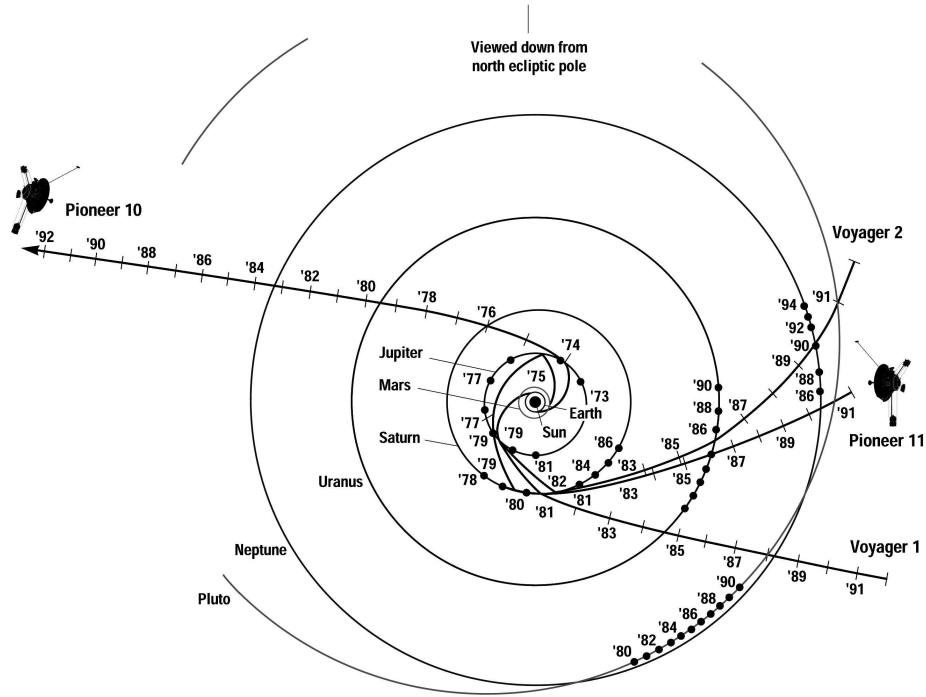,width=5in}
\end{center}
\caption
{\small
{Ecliptic pole view of Pioneer 10, Pioneer 11, and Voyager trajectories.  
Pioneer 11 is traveling approximately in the direction of the 
Sun's orbital motion about the galactic center.  The galactic center 
is approximately in the direction of the top of the figure.}} 
     \label{fig:pioneer_path}
\end{figure}

To illuminate how a drag force would yield an acceleration of the Pioneers towards the Sun, consider the dynamics of the situation.  
The Pioneers are on hyperbolic orbits, roughly in the plane of the 
ecliptic and parallel to the Sun's velocity vector in the galaxy.  
Pioneers 10 and 11 are moving in opposite directions with respect to the Sun, with Pioneer 10's velocity being opposite to the Sun's velocity vector.  That is, even though they are traveling away from the Sun on opposite sides of the solar system  (see Figure \ref{fig:pioneer_path}), to within the errors the anomalous accelerations of the two Pioneer craft are equal and are both directed towards the Sun.


The Pioneers are both moving at about 12 km/s ($\sim$ 2.5 AU/yr) relative to the Sun. Simultaneously, the local galactic rotation velocity is about 220 km/s with respect to the galactic origin and (as we return to below) the Sun is traveling at about 26 km/s relative to the local interstellar medium \cite{cox,intergas}. Therefore, if the anomaly is due to a drag force, the medium that is causing the drag must be, on average, locally ``radially at rest" (no relative radial momentum) about the Sun; for example as a sphere or a disk. This is true whether the medium is composed of normal matter or some unknown ``dark matter'' 
\cite{foot,foot2,dm1,dm2}.\footnote{Here we focus on ordinary matter 
as the possible origin of spacecraft drag.  
A dark matter hypothesis is discussed in Refs. \cite{foot,foot2}.} 


\section{The interplanetary medium}
\label{sec3}

The interplanetary medium is known to contain 
thinly scattered matter in the form of 
neutral hydrogen,  microscopic dust particles, and the hot solar-wind 
plasma of electrically charged particles (mainly protons and
electrons).  But the exact composition has long 
been debated, with many models put forward to describe  
the medium's nature and origin. As a result, 
limits on gas \cite{snow,scherer2000} and dust 
\cite{snow,scherer2000,teplitz1999} 
in the deep-space interplanetary medium are not precise, 
but the amount of gas is well known to be much less than the amount of 
dust.\footnote{
The gas is believed to come mainly from the interstellar medium as the
Sun revolves around the galaxy \cite{scherer2000}. 
It then has a velocity relative to the solar system of about 26 km/s 
\cite{cox,intergas}.  The gas drag velocity on a spacecraft is thus the vector sum of the craft's velocity and this 26 km/s.  The constant density of the gas is roughly equal to that of the solar wind at 20 AU, so only perhaps a few hydrogen atoms per 100 cm$^3$ \cite{scherer2000}.} 

As for dust, most of it should be in orbit about the Sun, even though some of the dust originates from the interstellar regime \cite{baguhl}-\cite{mann2}. Interstellar dust is distinguished from interplanetary dust {\it in situ} by its greater impact velocity on deep-space probes.  Starting with the Ulysses instruments, it has been measured.  It is found in \cite{mann1} that its total density is only 
\be
\rho_{\tt ISD} \lesssim 3 \times 10^{-26} \mathrm{g/cm}^3.
\label{ISD}
\ee
This careful determination, which we take as an upper bound, is compared to 
others in Table I of 
Ref. \cite{mann1}.\footnote{These other determinations are up to as much as an 
order of magnitude smaller.}  But in any event, and as will become clear below, 
the amount and relative velocity of the interstellar dust is much too small to have 
been seen as a drag on the Pioneers.  

The orbiting dust is, for us, mainly in  
the ``Kuiper belt,'' a disk-shaped region  extending 
roughly from the orbits of  Saturn and 
Neptune, $\sim$10-30 AU, out to about $\sim$80-120 AU from the Sun.  
It contains dust and many small icy bodies.\footnote{The Kuiper belt is 
now considered to be the source of short-period comets
whereas long-period comets are believed to be formed further away 
in the Oort cloud \cite{comet1}-\cite{comet4}.}
The dust will have a variety of eccentricities and inclinations, but 
for our simple purposes we can average out these different drag components 
into their circular velocity and have 
the drag velocity effectively be the radial velocity of the 
craft.\footnote{\label{dragvel}   
There is still a sideways drag caused by the average circular motion 
of the dust. But the side drag velocity is down from the radial velocity. 
It is lower by a factor $\sqrt{2}$ for a parabolic orbit and  down even more for escape hyperbolic orbits.} 

There currently is much effort devoted to understanding this region
\cite{teplitz1999}-\cite{mann2}, \cite{backman1995}-\cite{gruen}.
In particular, the study of the trans-Neptunian asteroids is a rapidly  
evolving field of research, with major observational and theoretical
advances in the last few years \cite{gruen}.  

The real problem is that, although measurements can determine the amount of interstellar dust, the same is not true for the interplanetary dust, which has 
a much lower relative velocity with respect to deep-space craft.  Further, 
starting with the Pioneers, the instruments that have been sent 
on missions to deep space have been sensitive only to varying-sized 
particles.\footnote{ 
For example, the Pioneer 10 and 11 impact detectors were sensitive to particles of masses $> 8 \times10^{-10}$ g and $> 6 \times10^{-9}$ g, respectively, at 
impact speeds of 20 km/s.  The Voyagers had plasma wave instruments that 
responded to impacts, but which were not calibrated for dust. 
(See, e.g., Ref. \cite{gruen2}.)} 
Instruments used have been combinations of  
mass spectrometers, dust impact detectors, plasma instruments, energetic particle analyzers, and magnetometers.  But 
even at their best the sensitivities of all the instruments 
on the deep-space craft so far launched have not been sufficient to detect all the individual effects of all the various mass and energy dust particles.  

The net result is that we 
are dependent on models for the interplanetary dust density, and these models vary greatly.  This is especially true for estimates of dust production in the Kuiper belt, which can vary by orders of magnitude \cite{mann2}.  A further complication is that the orbits of dust grains of different sizes will be most significantly influenced by different forces: gravitational, Poynting-Robertson drag, solar wind drag, and electromagnetic.  This makes understanding the {\it total} mass density even harder.   
 
A consensus view is that the {\it average} interplanetary dust density
may be almost two orders of magnitude larger (a factor of order 30) than the
interstellar dust density, and probably more \cite{mann1,kelsall}. 
Therefore, we can give a secure limit on the interplanetary dust 
density of 
\be
\rho_{\tt IPD} \gtrsim 10^{-24} \mathrm{g/cm}^3.
\label{IPD}
\ee
The lower bound $\rho_{\tt IPD}$ would yield $\sim 10^{21}$ g
in a 100 AU disk. This is not unreasonable since the younger, larger, Vega star 
is thought to have only approximately 8000 times this amount of dust in its disk 
\cite{irstellardust}. 

However, because of their relative velocities, all the particles 
discussed above will, as a matter of principle, produce a microscopic 
effective drag force on a passing spacecraft. Therefore, even though 
dedicated instruments may be optimized to yield the number density for  
particles of a certain 
mass, size, or kinetic energy, the drag on a large-area, low-mass spacecraft 
provides a way to ask about the total mass density distribution in g/cm$^3$.  

As we now come to, this fact allows {\it in situ, directly-measured} 
limits to be placed on the amount of deep-space interplanetary matter by 
using the Pioneer anomaly to bound overall the interplanetary mass density.


\section{Interplanetary density limits from the Pioneers}
\label{sec4}

Drag by the interplanetary medium on a spacecraft 
causes a deceleration of  \cite{piosail}
\be
a_{\tt s}(r)= -\mathcal{K}_{\tt s}~\frac{\rho(r)~v_{\tt s}^2(r)\,A_{\tt s}}{m_{\tt s}}, 
\label{dac}
\ee
where $\rho(r)$ is the density of the interplanetary medium, 
$\mathcal{K}_{\tt s}$ is the effective reflection/absorption/transmission coefficient of the craft for the 
particles hitting it, $v_{\tt s}(r)$ is the effective relative velocity of the craft with respect to the medium, $A_{\tt s}$ is the effective cross-sectional area of the craft, and $m_{\tt s}$ is its mass.  

In general $\mathcal{K}_{\tt s}$ is between 0 and 2.\footnote{
$\mathcal{K}_{\tt s}$ depends on the sizes and types of
the particles and especially on whether the particles are reflected 
($\mathcal{K}_{\tt s}=2$), 
absorbed ($=1$), or transmitted ($=0$) by the spacecraft.}  
Here we take $\mathcal{K}_{\tt s}$ to be a unit constant and the drag 
velocity to be $v_{\tt s} \sim 12$ km/s, the radial velocity of the 
Pioneers.\footnote{
The precise hyperbolic velocities of Pioneer 10 and 11 are about 12.2 and 11.6 km/s, respectively.  Also, the data discussed here is from about 
(see footnote \ref{footdistance}) 20-30 AU and 40-70 AU for Pioneers 10 and 11, respectively, and yielded slightly different values 
for the anomalous acceleration of the two craft.  (More details are in 
\cite{pioprd}.)  However, these differences cause effects that are within the systematic uncertainties of the result, so we use a common calculation for 
our one-significant-figure limits.}   
We can consider the effective area to be that of the Pioneers' 
antennae (radii of 1.37 m) and the mass (with half the 
fuel gone) to be 241 kg \cite{pioprd},\footnote{
Extending the observation in footnote \ref{dragvel}, the velocity of the 
Pioneers are 2 times larger than the orbital velocity at 25 AU, 
meaning a factor of 4 larger velocity effect for the radial vs. orbital drag.  Further, the effective area of the side of the Pioneer main bus and antenna is a factor $\sim 3$ smaller than the face of the Pioneer antenna. Therefore, we can ignore the side drag in this simple calculation.} or $(A/m)_{\tt P} = 0.245$ cm$^2$/g.
Given this, the critical unknown is $\rho(r)$. 

Below we will be considering densities of the form
\be
\rho_{\tt n}(r) = \rho_{\tt 0n} \left(\frac{r_{\tt 0}}{r}\right)^n,
\ee
where the $\rho_{\tt 0n}$ and $r_{\tt 0}$ are constants, with $r_{\tt 0}$ set 
to be at the beginning of the Pioneer data interval, 20 AU.


\subsection{Uniform density}

By assuming that the Pioneers' {\it entire} anomalous acceleration is due to a drag force, we can calculate that, at distances from 20 to 70 AU from the Sun,  an 
axially-symmetric dust distribution with a {\it constant}, uniform density 
\be
\rho_{\tt P}(r) \lesssim \rho_{\tt 00} = 3 \times 10^{-19}~\mathrm{g/cm}^3     
\label{rhoP}
\ee
could have produced the {\it constant} anomaly.  Eq.~(\ref{rhoP})
places an {\it in situ, experimental} limit on the density of 
the interplanetary medium, even though it is larger by a factor of 300,000
than $\rho_{\tt IPD}$ of Eq.~(\ref{IPD}), the latter a number more like those  
usually thought of for deep space.  Indeed, the limit of Eq.~(\ref{rhoP}) 
corresponds to about 200,000 atomic-masses/cm$^3$.

Ruling against this upper bound being, in fact, a measure of the density is that a drag acceleration from Kuiper belt dust should not have been constant across the Pioneer data range. Not only should there be boundaries from bands of dust, but concentrations of ``Kuiper-Belt Objects'' (KBO) at 39.4 AU and 47.8 AU, corresponding to Neptune resonances of 3:2 and 2:1 \cite{malhotra1995,malhotra1996}, have been discovered.  The KBO concentrations will at least affect dust creation from collisions.  But questions remain on exactly how the steady-state mass density of the dust will be affected both by the concentrations themselves and also by the resonances that created them. 
These density spatial variations are widely discussed in the literature 
\cite{snow,scherer2000,mann1,mann2,backman1995,stern1996,kelsall,gruen}.\footnote{A related consideration is if the mass in the Kuiper belt could produce a {\it gravitational acceleration} that causes the Pioneer anomaly. In Section VII.E and Figure 15 of \cite{pioprd} it is shown that even generous  
($\sim 5 \times 10^{-18}~\mathrm{g/cm}^3$),
although reasonable, models of Kuiper belt densities can not produce the Pioneer anomaly by three and more orders of magnitude.}

Therefore, a drag should have shown an increasing effect as the 
spacecraft approached into belts or concentrations and a decreasing 
effect as it receded from belts and concentrations, even with an 
approximate uniform density within the overall belt.\footnote{ 
One could also argue that what one is seeing in the anomaly is evidence 
for dark matter causing a drag \cite{foot,foot2}.
But even then one has to explain why this matter is a constant density 
in the regime penetrated by the Pioneers.} 
However, independent of the lack of a spatial variation in the 
anomaly, which implies any actual  
average density that exists is lower than $\rho_{\tt P}$, 
Eq.~(\ref{rhoP}) remains a {\it model-independent, in situ, experimental} upper bound for the interplanetary density. 


\subsection{Density varying as $\mathbf{1/r}$}

Even lower bounds can be placed if one {\it presumes} a model density that varies with distance.  For example, consider a density that varies as 
\be
\rho_{\tt 1/r}(r) \sim \rho_{\tt 01}\left(\frac{r_0}{r}\right).
\ee
Conservatively it is limited by 
the lack of variation seen in the anomaly.  The main part of the anomaly 
then must be due to another origin, but the size of the anomaly's 
total error,\footnote{The total error is dominated by systematics, many of which are constant or nearly so \cite{pioprd}.  Therefore, one could argue that the numbers in Eqs.~(\ref{rho/r1}) and (\ref{rho/r2}) below can be reduced accordingly.} $\sigma_{\tt P} = 1.33 \times 10^{-8}$  cm/s$^2$, 
places a limit on how much matter there is.  Between 20 and 70 AU the $1/r$ fall off of the density would imply a change in the acceleration of 
$\sigma_{\tt P}$.  This yields a value for $\rho_{\tt 01}$ and hence the result 
\be
\rho_{\tt 1/r}(r) \lesssim 6 \times 10^{-20}
\left(\frac{20~\mathrm{AU}}{r}\right) ~\mathrm{g/cm}^3, ~~~~~~~~
    20~\mathrm{AU} \le r \le 70~\mathrm{AU}.   
\label{rho/r1}
\ee


\subsection{Isothermal density}

By the same argument as above, an isothermal  density varying as 
\be
\rho_{\tt isoth}(r) \sim \rho_{\tt 02}\left(\frac{r_0}{r}\right)^2
\ee
yields a limit 
\be
\rho_{\tt isoth}(r)  \lesssim 5 \times 10^{-20}
\left(\frac{20~\mathrm{AU}}{r}\right)^2 ~\mathrm{g/cm}^3, ~~~~~~~~
    20~\mathrm{AU} \le r \le 70~\mathrm{AU}.   
\label{rho/r2}
\ee

However, there is a caveat with this model.  
It proposes that $\sigma_{\tt P}$ provides a bound on the size of the variation
of the unmodeled non-gravitational drag force between 20 and 70 AU.  At 20 AU it  
is of size $1.33 \times 10^{-8}$  cm/s$^2$ and it falls off as the square of the distance from there. 
But at 20 AU there is another non-gravitational force that falls off as the 
square of the distance, is of similar size 
($\sim 5 \times 10^{-8}$  cm/s$^2$),
but of opposite sign.  It is produced by the solar radiation pressure from the Sun on the spacecraft \cite{pioprd}. Therefore, to distinguish this type of drag force from radiation pressure would entail extremely precise modeling and orbit determination. 

Also, note that the final two densities above,  
$\rho_{\tt 1/r}(r)$ and $\rho_{\tt isoth}(r)$,   
must cut off closer in to the Sun. Otherwise they would produce too large a drag.

\begin{figure}[h!]
\begin{center}\noindent    
\epsfig{figure=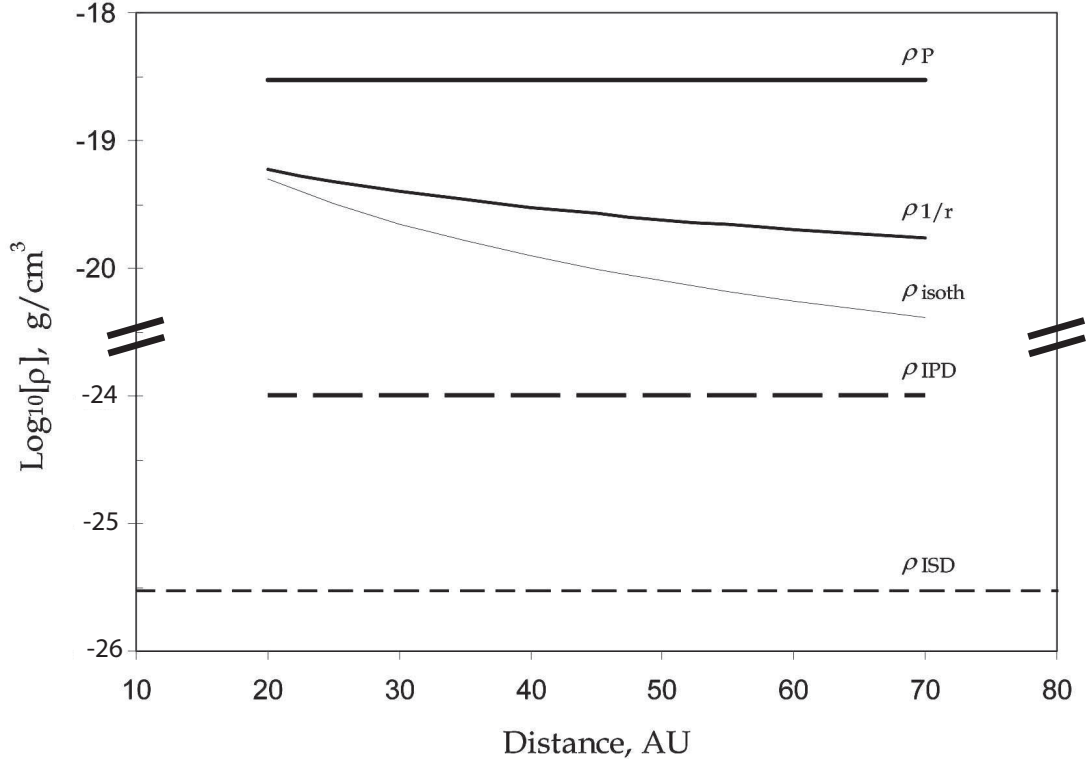,width=145mm}
\end{center} \vskip -15pt
\caption
{\small
{Plots of the log-to-the-base-10 of density (in g/cm$^3$) vs distance (in AU), for (from top to bottom)
the uniform density limit, $\rho_{\tt P}(r)$, 
the $1/r$ model density limit, $\rho_{\tt 1/r}(r)$, 
the isothermal model density limit, $\rho_{\tt isoth}(r)$,
the interplanetary dust model estimate, $\rho_{\tt IPD}$, and 
the interstellar dust estimation, $\rho_{\tt ISD}$.}} 
     \label{fig:densitymodels}
\end{figure}

In Figure \ref{fig:densitymodels} we show the above three bounds, $\rho_{\tt P}(r)$,  $\rho_{\tt 1/r}(r)$, and $\rho_{\tt isoth}(r)$, as well as the estimates for interstellar and interplanetary dust, $\rho_{\tt ISD}$ and $\rho_{\tt IPD}$, quoted in Eqs.~(\ref{ISD})-(\ref{IPD}).


\section{Conclusions and future considerations}
\label{sec5}

In this paper we analyzed the possibility that the Pioneer anomaly is the result of a drag force from dust distributed in the outer solar system.   Our analysis showed that for this mechanism to work, one would need the presence of 
dust with a density on the order of
$\sim$$(5 - 30) \times 10^{-20}$ g/cm$^3$, 
in the region 20 to 70 AU from the Sun. 
This is unexpectedly high.   
Our present knowledge of dust formation processes in the outer regions of 
the solar system implies that such a high density is not realistic.  
Even so, the accuracy of the Pioneer orbit reconstruction allowed us to 
place both model-independent and model-dependent, {\it in-situ}, experimental limits  on the mass density of dust in the outer solar system.  

The results presented in this paper can also be used to help further motivate a mission to explore the origins and evolution of our solar system.  Of course, among the  objectives of such a mission would be to precisely map the gas and dust distributions in the solar system at various heliocentric distances and latitudes.
Gas and dust detectors are typically among the standard set of instruments in deep-space missions, whatever their objectives. 

However, we emphasize that any {\it low-mass}, deep-space 
mission with precise radio-science experiments,
like a mission that would attempt to explore the 
Pioneer anomaly \cite{piofind},\cite{mission}-\cite{cosmic}, 
should be prepared to ascertain if any observed effect is due to a drag force, 
thereby providing an independent limit (or even measurement) of the matter density, 
especially in the outer regions of the solar system.

Eq.~(\ref{dac}) shows that the drag acceleration scales as 
$(A_{\tt s}~v_{\tt s}^2/m_{\tt s})$.   Comparing this quantity for the 
Pioneers to that for any other spacecraft quickly yields a figure of merit 
on the ability of this other spacecraft to obtain a better limit on the interplanetary density (or possibly even a measurement of it). 
Another spacecraft could also look for an 
indication of a drag force by using three-dimensional navigation to determine 
if any anomalous force were directed 
along the velocity vector of the craft instead of along the vectors towards the Sun, the Earth of along the spin axis. (This point is explained in detail in \cite{piofind}.) 

As an example of such a mission, 
consider a spacecraft design whose architecture is symmetric in the forward and backward directions, having two oppositely 
facing antennae \cite{piofind,cosmic}.  Its cross section and 
area are similar to the Pioneers, but its measured effect would be 
boosted by the square of its (presumed) larger velocity.

A second example is a formation flying concept
\cite{enigma,cosmic}, where the position of a 5 cm radius 
ball covered with corner-cubes weighing 5 kg is tracked from a mother ship (which in turn is tracked from Earth).  The (area/mass) of this 
concept is 16 times smaller than that of the Pioneers.  Therefore, to 
measure the same drag acceleration the formation flying mission would 
need to be traveling 4 times faster than the 
Pioneers.\footnote{The mother ship (satellite) would probably have an
$(A_{\tt s}/m_{\tt s})$ similar to that of the Pioneers.  But in the
current conception \cite{enigma,cosmic} the main point
is to eliminate systematics from the sub-satellite ball,
the satellite itself not being optimized to have low systematics.
This means the two craft would drift apart.
Even for a difference in acceleration between the two craft of
only $a_P$, in one month they would separate by 3 km, this distance
increasing quadratically with time after that.
To maintain laser tracking of the subsatellite, the distance change
would periodically have to be compensated for by thrust maneuvers on
the mother satellite.} 

Both of the above concepts are designed to reduce systematic errors by two to three orders of magnitude.  This would allow even better limits to be placed on $r$-dependent densities.  Further, the same idea could be used on a solar-sail mission to deep space 
{\it if} the sail were {\it not} jettisoned past Jupiter's orbit \cite{piosail}, as is usually conceived of.   

Lastly, if one assumes a drag force is causing the Pioneer anomaly, then a test should also try to determine if the matter starts at about 10 AU.\footnote{Such a Kuiper belt model starting at 10 AU has been suggested \cite{10AU}.} This is where the early, not yet precisely analyzed, Pioneer 11 data seems to show the anomaly ``turning on'' as the craft passed by Saturn, turned radially outward, and reached its hyperbolic escape velocity of 11.6 km/s   
\cite{pioprl,pioprd,mex,cospar_04}. 

The exact architecture for a mission to explore the Pioneer anomaly, including the spacecraft and mission designs, the set of critical instruments, and the launch options, is currently being investigated \cite{piofind},\cite{mission}-\cite{cosmic}.  We await progress on this front.


\section*{Acknowledgments}

We thank Ingrid Mann and Klaus Scherer for valuable comments and information 
on the interplanetary medium, and Nicole Amecke and Hansjoerg Dittus 
for corrections and suggestions for the manuscript.  
M.M.N. acknowledges support  by the U.S. DOE.     
The work of S.G.T. and J.D.A. was performed at the 
Jet Propulsion Laboratory, California Institute of Technology, under
contract with the National Aeronautics and Space Administration.  




\begin{thebibliography}{99}


\bibitem{pioprl} J. D. Anderson, P. A. Laing, E. L. Lau, A. S. Liu,
M. M. Nieto,  and S. G. Turyshev,  
{ Phys. Rev. Lett.} 81 (1998) 2858,
gr-qc/9808081.

\bibitem{pioprd} J. D. Anderson, P. A. Laing, E. L. Lau, A. S. Liu,
M. M. Nieto, and S. G. Turyshev, 
{ Phys. Rev. D.} 65 (2002) 082004,
gr-qc/0104064.

\bibitem{markward} C. Markward, 
eprint gr-qc/0208046.
 
\bibitem{courtens} N. Didon, J. Perchoux, and E. Courtens, 
Universit\'e de Montpellier report (1999). 

\bibitem{foot} R. Foot and R. R. Volkas, 
{ Phys. Lett. B} 517 (2001) 13, 
gr-qc/0108051.

\bibitem{foot2} R. Foot, 
Int. J. Mod. Phys. A 19 (2004) 3807, 
astro-ph/0309330.

\bibitem{piofind} M. M. Nieto and S. G. Turyshev, 
Class. Quant. Grav. 21 (2004) 4005,
gr-qc/0308017.

\bibitem{newplanets}  S. Seager, 
Earth and Planet. Sci. Lett. 208 (2003) 113.

\bibitem{irstellardust} H. J. Walker, 
Icarus 143 (2004) 147.



\bibitem{cox} A. N. Cox, {\it Allen's Astrophysical Quantities}, 4th ed. 
(Springer, New York, 2000), p. 569, 494. 

\bibitem{intergas} R. Lallement, Adv. Space Res. 13 (1993) 113.

\bibitem{dm1}  A. B. Lahanas, N. E. Mavromatos, and D. V. Nanopoulos, 
Int. J. Mod. Phys. D 12 (2003) 1529.

\bibitem{dm2}  J. M. Overduin and P. S. Wesson, 
Phys. Rep. 402 (2004) 267.

\bibitem{snow} T. P. Snow, J. Geophys. Res. A 105 (2000) 10,239.

\bibitem{scherer2000} K. Scherer, 
J. Geophys. Res. A 105 (2000) 10,329.

\bibitem{teplitz1999} V. L. Teplitz, S. A. Stern, J. D. Anderson, D.
Rosenbaum, R. J. Scalise, and P. Wentzler, 
{ Astrophys. J.} 516 (1999) 425,
astro-ph/9807207.

\bibitem{baguhl}  M. Baguhl, E. Gr\"un, D. P. Hamilton, G. Linkert, 
and P. Staubach, Space. Sci. Rev. 72 (1995) 471.

\bibitem{mann1} I. Mann and H. Kimura, 
J. Geophys. Res. A 105 (2000) 10,317.

\bibitem{mann1.5}
R. Mukai, A. Higuchi, P. S. Lykawka, H. Kimura, I. Mann, and S. Yamamoto, 
Adv. Space Res.  34 (2004) 172. 

\bibitem{mann2} I. Mann, A. Czechowski, and S. Grezdzielski,
Adv. Space Res.  34 (2004) 179. 

\bibitem{comet1} P. Wiegert and S. Tremain,
Icarus 132 (1999) 84.

\bibitem{comet2}  M. E. Bailey, 
Science 206 (2002) 2151.

\bibitem{comet3} J. D. Anderson, S. Turyshev, and M. M. Nieto, 
Bull. Am. Astron. Soc., 34 (2002) 1172.

\bibitem{comet4} D. P. Whitmire and J. J. Matese, 
Icarus 165 (2003) 219.

\bibitem{backman1995} G. E. Backman, A. Dasgupta, and R. E. Stencel,  
{ Astrophys. J.} {450} (1995) L35.

\bibitem{stern1996}  S. A. Stern, 
{Astron. Astrophys.}  {310} (1996) 999.

\bibitem{jdapio98icarus} J. D. Anderson, E. L. Lau, K. Scherer, D. C. Rosenbaum,
and V. L. Teplitz, Icarus 131 (1998) 167.

\bibitem{kelsall} T. Kelsall, J. L. Weiland, B. A. Franz, W. T. Reach, 
R. G. Arendt, E. Dwek, H. T. Freudenreich, M. G. Hauser, S. H. Moseley, 
N. P. Odegard, R. F. Silverberg, and E. L. Wright, 
Astrophs. J. 508 (1998) 44. 

\bibitem{gruen}  E. Gr\"un, B. {\AA}. S. Gustafson, S. F. Dermott, and H. Fechtig, eds., {\it Interplanetary Dust} (Springer, Berlin, 2001).

\bibitem{gruen2} E. Gr\"un, M. Baguhl, H. Svedhem, and H. A. Zook, 
in: \cite{gruen} p. 295.


\bibitem{piosail} M. M. Nieto and S. G. Turyshev, 
Int. J. Mod. Phys. D {13} (2004) 899, 
astro-ph/0308108.

\bibitem{malhotra1995} R. Malhotra, 
{ Astron. J.} 110 (1995) 420.

\bibitem{malhotra1996} R. Malhotra,  
{ Astron. J.} {111} (1996) 504.

\bibitem{mission} J. D. Anderson, S. G. Turyshev, and M. M. Nieto, 
{ Int. J. Mod. Phys. D.}
{11} (2000) 1545, 
gr-qc/0205059.

\bibitem{orfeu} O. Bertolami and M. Tajmar, ``Gravity Control and Possible Influence on Space Propulsion: A Scientific Study,'' ESA Report CR (P)
4365 (2002).

\bibitem{enigma} U. Johann and R. F\"orstner, ``Enigma,'' EADS
Astrium GmbH Pioneer proposal to ESA (2004).  

\bibitem{pluto}  A. Rathke, in: {\it Frontier Science 2004: Physics
and Astrophysics in Space} (to be published), astro-ph/0409373.

\bibitem{cosmic}
H. Dittus, C. L\"ammerzahl, S. Theil, B. Dachwald, W. Seboldt, 
W. Ertmer, E. Rasel, R. Foerstner, U. Johann, F. W. Hehl, C. Kiefer, 
H.-J. Blome, R. Bingham, B. Kent, T. J. Sumner, 
O. Bertolami,  J. L. Rosales, B. Christophe, B. Foulon, P. Touboul, 
P. Bouyer, S. Reynaud, C. J. de Matos, 
C. Erd, J. C. Grenouilleau, D. Izzo, A. Rathke, 
J. D. Anderson, S. W. Asmar, S. G. Turyshev, 
M. M. Nieto, and B. Mashhoon, {\it A Consolidated Cosmic Vision Theme 
Proposal to Explore the Pioneer Anomaly,} submitted to ESA (Oct. 2004).

\bibitem{10AU} 
N. N. Gorkavyi, L. M. Ozernoy, T. Taidakova, and J. C. Mather,
astro-ph/0006435.

\bibitem{mex} M. M. Nieto, S. G. Turyshev, and  J. D. Anderson,
in:  {\it Proceedings of the 2nd Mexican Meeting on Mathematical and
Experimental Physics}, eds: A. Mac\'ias, C. L\"ammerzahl, and D. Nu\~nez (American
Institute of Physics, NY, to be published), gr-qc/0411077.

\bibitem{cospar_04}  S. G. Turyshev, M. M. Nieto,  and J. D. Anderson, 
Adv. Space Res. (to be published, 2005), 
{\it Proc. 35th COSPAR Sci. Assembly} (Paris, 2004), 
gr-qc/0409117.


\end{thebibliography}
\end{document}